# Creating Learning Scaffolds for Engineering Design Using Concept Catalyst

Madhuri Singh, Gennie Mansi, Mark Owen Riedl
msingh365@gatech.edu, gennie.mansi@gatech.edu, riedl@gatech.edu
Georgia Institute of Technology

**Abstract:** K-12 teachers employ Engineering Design Challenges to help students learn about the Engineering Design Process hands-on. They use techniques like hard scaffolding questions to guide the students as they think through the different stages of the engineering design process. While useful, the creation of these questions adds to the teacher's preparation time for their classes. Concept Catalyst uses Large Language Models to assist teachers with the rapid creation of scaffold questions for engineering design challenges. Unlike open-ended chat, Concept Catalyst uses LLMs to summarize and decompose an engineering design challenge into the concepts that students will engage with, allow the teacher to visually manipulate and link related concepts, and to propose scaffolding questions for the teacher to modify or accept.

## Introduction

K-12 engineering teachers help students learn about the Engineering Design Process (EDP) through design challenges, such as constructing a bridge or catapult. Engineering design challenges require students to break down a problem, think critically about it, develop an approach to address it, then iteratively implement, evaluate and redesign their approach. Writing documentation is a key practice used to teach the EDP—tables, graphs, diagrams, and other written references help students reflect on discussions, trade-offs between designs, and prototype failures (Mangold & Robinson, 2013).

To help students learn how to use documentation to iterate on their project solutions, K-12 engineering teachers use *scaffolding questions* (Douglas & Chiu, 2012). Soft scaffolding involves teachers dynamically offering support as they monitor students, while hard scaffolding involves anticipating and addressing students' needs ahead of time (Saye & Brush, 2002).

To create learning scaffolds, teachers draw on various teaching and professional experiences (Nadelson et al., 2014) to create and contextualize new curriculum and projects (Chiu et al., 2021). Reflective practices—in which teachers critically examine their teaching experiences and knowledge—are central to K-12 engineering teachers' iteration on classroom pedagogy. Reflective practices can improve teachers' understanding of engineering content (Radhakrishnan et al., 2021), discover strengths and weaknesses of their teaching (Mesutoglu & Baran, 2021), and revise lesson plans (Mesutoglu & Baran, 2021). However, creating scaffolds is both time intensive and challenging. Teachers must anticipate potential difficulties that students may face and tailor their curriculum to students' needs (Saye & Brush, 2002), adding to an intensive workload (Creagh et al., 2025).

Generative AI technologies hold the potential to support teacher activities, including lesson planning and assessment creation, thus decreasing teacher workload (Kasneci et al., 2023; Ravi et al., 2025). Teachers have expressed openness to using LLM-based tools designed to support project-based learning tasks (Ravi et al., 2025). While LLMs may not be reliable enough in a fully automated capacity (Schneider et al., 2025), they can be effectively used with teachers' oversight (Nagy et al., 2023), hence creating a collaborative system of teachers with AI that supports teacher agency and benefits both students and the teachers (Li et al., 2025; Yan et al., 2025).

In this paper, we introduce an AI-based tool, Concept Catalyst, which helps teachers think through engineering design challenges using AI summarization, visualizations, and the synthesis of scaffolding questions once they have decided on which aspects of the problem to focus. Concept Catalyst is focused on helping teachers, not replacing them. The tool's design is based on prior work which introduced the framework to K-12 Engineering teachers in a Wizard-of-Oz study in which teachers reported positive feedback on the efficiency, quality, and diversity of content they wrote, explaining how it helped them more deeply engage with the EDP.

## Concept catalyst

Concept Catalyst is a web-based tool that assists teachers in analyzing a design challenge and creating scaffolding questions for students. Concept Catalyst prompts teachers to highlight important concepts in paragraph summary of a design challenge, and then visually organize the concepts based on how they connect. Finally, teachers work with Concept Catalyst to synthesize scaffolding questions for their students.

**Figure 1**

*A screenshot of the SUMMARIZE step*

Concept Catalyst

This is an LLM-assisted tool that helps you breakdown EDP problems into germane concepts, visualize their connections and generate relevant questionnaires to serve as scaffolds for the students.

A short description of the process to follow:
1. In this page you will select the EDP problem that you want to work on. The tool will generate a 500 word summary of the selected problem.
2. In the next page, you will select the concepts from the problem summary and see how they connect with each other.
3. You will select groups of concepts for which you want the questions to be generated, and the LLM will create the questions for you.
4. Finally, you select the questions you like and print them for your usage.

To begin, please select how you would like to enter the EDP problem summary

| | |
|---|---|
| File Input | Please upload a file containing a summary of the EDP problem: Select file Supported file types : .pdf, .doc, .docx |
| Text Input | Please enter a summary of the EDP problem here: Continue with this summary |

The goal of Concept Catalyst is to reduce the amount of time taken in generating scaffolding questions. Concept Catalyst uses visualizations to support teachers' reflective practices with their curriculum while enabling them to easily leverage LLMs to generate scaffolding questions without the open-endedness of chat interfaces. Concept Catalyst avoids the issue of opacity of LLM-based decision making by keeping the decision making task entirely with the teacher, and employing the LLM as a generator of suggestions. However, we acknowledge that usage of this tool might still face structural issues, like the access of AI-supported tools being disproportionate between different socio-economic classes (Yan et al., 2025).

Concept Catalyst focuses teachers' attention on the Engineering Design Process (EDP) while keeping the LLM hidden behind an interface that guides teachers to create a concept map. Concept maps, knowledge graphs, and tree diagrams are all structured knowledge representations that can support the integration of new information with prior knowledge (Moreno & Mayer, 2007), a key aspect of reflection shown to help improve teacher practices (Ardito, 2022). Concept Catalyst uses teachers' concept maps to suggest scaffolding questions, which teachers can edit. Concept Catalyst is entirely teacher-facing and only outputs scaffold questions that can be printed and shared with students, who do not interact with the system.

Critically, Concept Catalyst places teachers in control of the scaffolds' content—helping teachers leverage the benefits of generative AI without removing their agency. Below, we detail how teachers work with Concept Catalyst through three stages: (1) Summarize, (2) Conceptualize and (3) Synthesize.

## Summarize

In the SUMMARIZE step (see Figure 1) the teacher is prompted to input a paragraph summary of the design challenge or upload a unit map for the design challenge. If the teacher uploads a unit map, the tool processes it and displays a 200-word summary. Regardless, they can choose to modify it. Once the teacher is satisfied, they can move ahead by clicking on a 'Continue' button. See Figure 1 for a screenshot of this step.

**Figure 2**
*A screenshot of the CONCEPTUALIZE step*

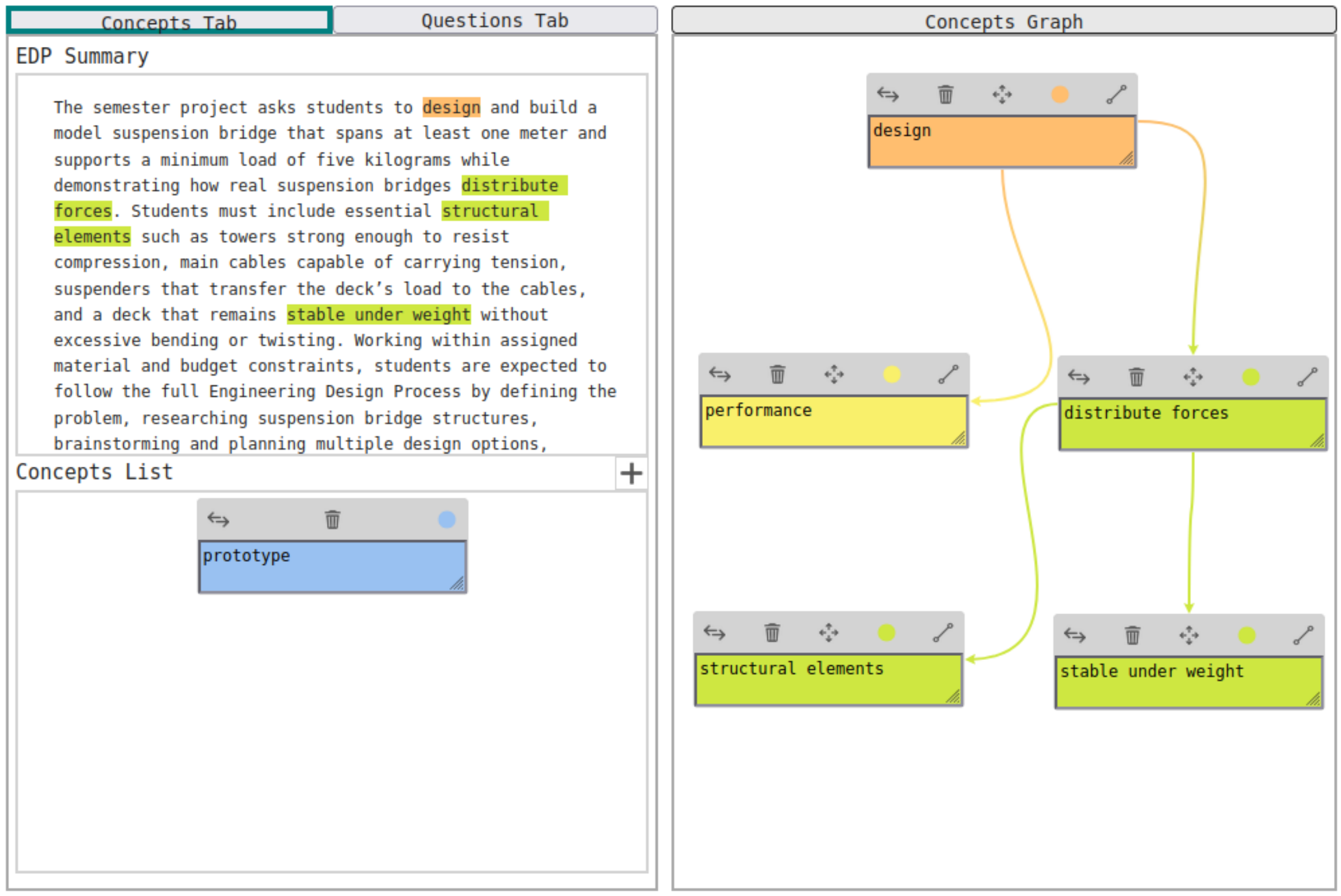

## Conceptualize

In the CONCEPTUALIZE step, teachers are prompted to highlight key concepts that students should address in their design challenge and then visually connect the concepts that are related. The left half of the screen, titled 'Concepts Tab', holds two sections. The section on the top displays the summary that the teacher just approved. The bottom section, empty in the beginning, is a waiting area for the newly created concepts. The right half of the screen, titled 'Concepts Graph', is dedicated to a single section where teachers can move the concepts that they want to explore and connect them with lines.

The teacher begins by using their mouse to highlight any words or phrases in the summary that depict pertinent concepts. Every such highlight generates a concept-button in the waiting area. The teacher can also choose to create custom concepts by just clicking on the '+' button. The teacher then moves the concept-button from the waiting area (lower left) to the graph area. The graph area provides two functionalities to the teacher: freely moving the concept-buttons around, and connecting them via lines. The teacher can explore the relationships between different concepts by moving the respective buttons closer or farther, upward, or downward, or by connecting them in a graph.

This visual exploration is aimed at providing the teacher with an avenue to apply their creativity and finding interesting concepts that inform the problem and the ways that they interplay.

The teacher can switch between these sections as many times as they need. Once they feel comfortable with the concept-graph, they can choose to switch to the Questions tab. See Figure 2 for a screenshot depicting the *CONCEPTUALIZE* step.

## Synthesize

The SYNTHESIZE step is where the scaffolding questions are generated. This is the third screen of the tool, which shares its right half with the previous screen where the 'Concepts Graph' stays as it was left in the previous screen, while the left half of the screen, titled 'Questions tab', hosts question groups. The teacher attaches any number of concepts with a given question group by just clicking them. Each associated concept is previewed in the question group. Once the teacher deems the concept sufficient for a question group, they click the respective 'Generate questions' button. The tool generates five questions addressing the concepts that a student should consider while working on their project.

**Figure 3**
*A screenshot of the SYNTHESIZE step*

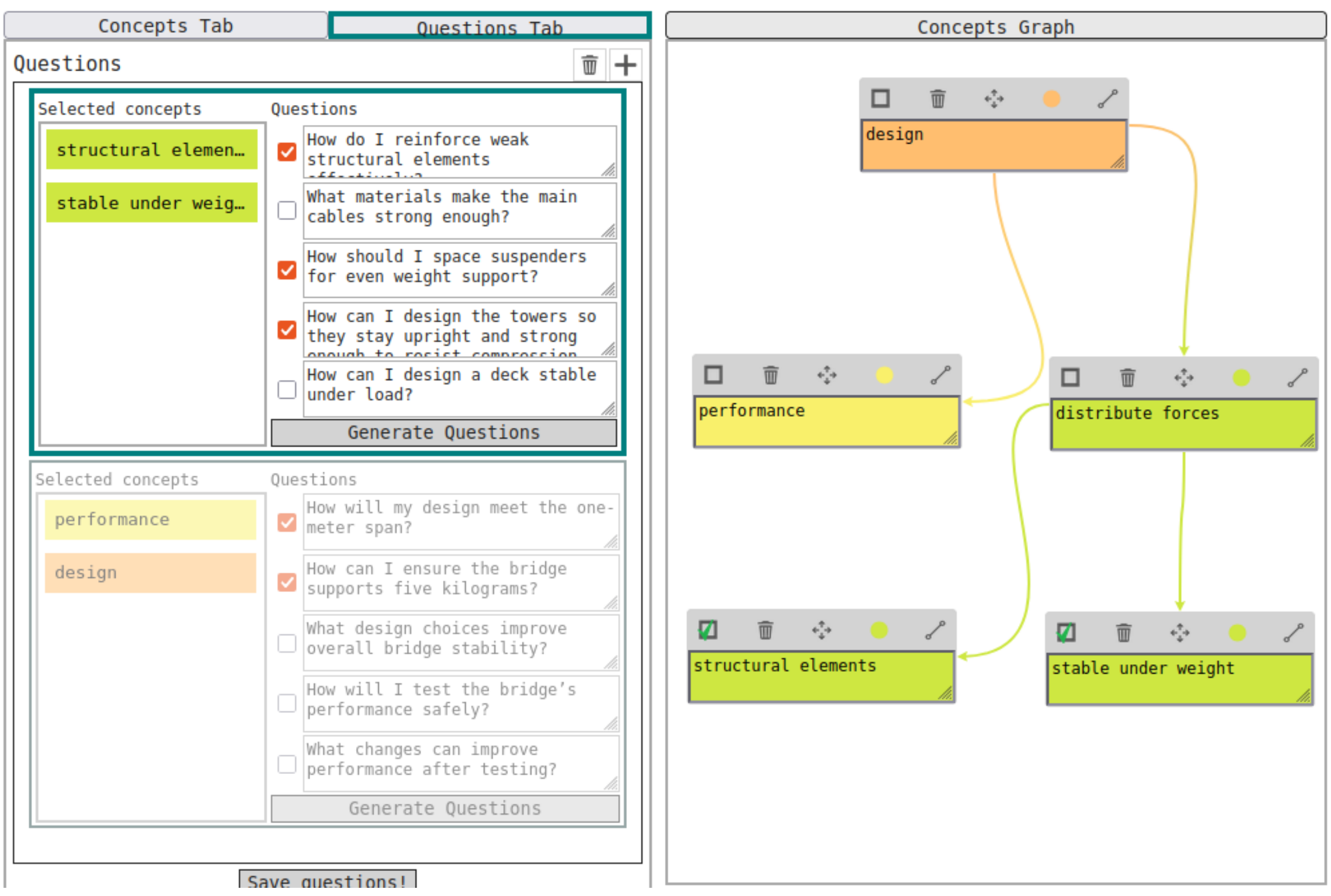

The teacher can generate any number of question groups with any combination of concepts they need. The teacher has complete freedom to accept, reject or modify the generated questions. The accepted questions are added to a question bank. Once the teacher considers their work complete, they can end the process by clicking the 'Save questions' button which will give them a final preview of the approved questions and then print them to paper or a pdf document. See Figure 3 for a screenshot of this step.

## Conclusions

Concept Catalyst is designed to reduce teachers' cognitive load while generating scaffolding questions for design challenges. At every single point of this interaction, the teacher can choose to accept, reject, or modify any artifacts created by the tool. The three stages—Summarize, Conceptualize and Synthesize—are not linear, allowing for an iterative flow that supports teachers' reflection of the problem as visualization of the concepts and their relationships gradually add clarity and suggest further possibilities.

## References

Alfarwan, A. (2025). Generative AI use in K-12 education: a systematic review. Frontiers in Education, 10.

Ardito, G.P. (2022). Concept Mapping: A Tool for Adolescent Science Teachers to Improve Learning Activity Design. In: Rezaei, N. (eds) Integrated Education and Learning. Integrated Science, vol 13. Springer.

Chiu, J., Fick, S., McElhaney, K., Alozie, N., & Fujii, R. (2021). Elementary Teacher Adaptations to Engineering Curricula to Leverage Student and Community Resources. Journal of Pre-College Engineering Education Research, 11(05).

Creagh, S., Thompson, G., Mockler, N., Stacey, M., & Hogan, A. (2025). Workload, work intensification and time poverty for teachers and school leaders: a systematic research synthesis. *Educational Review*, *77*(2), 661–680.

Douglas, E.P. and Chiu, C. (2012). Process-oriented Guided Inquiry Learning in Engineering. Social and Behavioral Sciences, 56(253–257).

Kasneci, E., Sessler, K., Küchemann, S., Bannert, M., Dementieva, D., Fischer, F., Gasser, U., Groh, G., Günnemann, S., Hüllermeier, E., Krusche, S., Kutyniok, G., Michaeli, T., Nerdel, C., Pfeffer, J., Poquet, O., Sailer, M., Schmidt, A., Seidel, T., Stadler, M., Weller, J., Kuhn, J., and Kasneci, G. (2023). ChatGPT for good? On opportunities and challenges of large language models for education. Learning and Individual Differences, 103 (102274).

Kokoç, M. (2024). Factors influencing K-12 teachers' experiences of using Generative AI Tools: opportunities and barriers. Journal of E-Learning and Knowledge Society, Vol 20 No 3(2024), Resources and Challenges in the Context of ICT".

Li, M., Wang, Y., & Han, X. (2025). Empowering Teacher Agency in the Era of Artificial Intelligence: Challenges and Strategies. Lecture Notes in Computer Science, 3–16.

Mangold, J., & Robinson, S. (2013). The engineering design process as a problem solving and learning tool in K-12 classrooms. ASEE Annual Conference and Exposition.

Mesutoglu, C., & Baran, E. (2020). Integration of engineering into K-12 education: a systematic review of teacher professional development programs. Research in Science & Technological Education, 39(3).

Moreno, R., & Mayer, R. (2007). Interactive Multimodal Learning Environments. Educational Psychology Review 19.

Nadelson, L., Seifert, A.L., & Mckinney, M. (2014). Place-based stem: Leveraging local resources to engage K-12 teachers in teaching integrated stem and for addressing the local stem pipeline. Proceedings of the ASEE Annual Conference & Exposition.

Nagy, S., McInnes, R., & Airey, L. (2023). GEN-AI: A TRANSFORMATIVE PARTNER IN COLLABORATIVE COURSEDEVELOPMENT. International Journal on Innovations in Online Education, 7(2), 57–73.

Radhakrishnan, D., Capobianco, B. M., & DeBoer, J. (2021). Kenyan engineering teachers building reflective practice. *Reflective Practice*, *22*(5), 697–711.

Saye, J.W., Brush, T. Scaffolding critical reasoning about history and social issues in multimedia-supported learning environments. *ETR&D* **50**, 77–96 (2002).

Ravi, P., Masla, J., Kakoti, G., Lin, G., Anderson, E., Taylor, M., Ostrowski, A., Breazeal, C., Klopfer, E., & Abelson, H. (2025). Co-designing Large Language Model Tools for Project-Based Learning with K12 Educators. ArXiv (Cornell University).

Schneider, J., Schenk, B., & Niklaus, C. (2024). Towards LLM-Based Autograding for Short Textual Answers. In Proceedings of the 16th International Conference on Computer Supported Education (pp. 280–288). 16th International Conference on Computer Supported Education. SCITEPRESS - Science and Technology Publications.

Yan, L., Suleman Abdullah Alwabel, A., & Mohamad, U. H. (2025). AI-Powered Education: Transforming Teacher-Student Interactions and Advancing Sustainable Learning Practices. European Journal of Education, 60(4). Yang, S., Trainin, G., & Appleget, C. (2024). Teacher Use of Generative AI for Read-Aloud Question Prompts. The Reading Teacher, 78(4), 230–235.

## Acknowledgments

This material is based upon work supported by the National Science Foundation under Grant No. IIS-2119135 . Any opinion, findings, and conclusions or recommendations expressed in this material are those of the authors(s) and do not necessarily reflect the views of the National Science Foundation.